\documentclass[conference]{IEEEtran}
\IEEEoverridecommandlockouts

\usepackage{cite}
\usepackage{amsmath,amssymb,amsfonts}
\usepackage{algorithmic}
\usepackage{graphicx}
\usepackage{textcomp}
\usepackage{xcolor}

\usepackage{booktabs}
\usepackage{multirow}
\def\BibTeX{{\rm B\kern-.05em{\sc i\kern-.025em b}\kern-.08em
    T\kern-.1667em\lower.7ex\hbox{E}\kern-.125emX}}
\begin{document}
\title{Effective Integration of KAN for Keyword Spotting
}

\author{Anfeng Xu$^{\dagger}$$^{\star}$, Biqiao Zhang$^{\ddagger}$,
 Shuyu Kong$^{\ddagger}$, Yiteng Huang$^{\ddagger}$,
 Zhaojun Yang$^{\ddagger}$, Sangeeta Srivastava$^{\ddagger}$, Ming Sun$^{\ddagger}$ \\ \\

\thanks{$^{\star}$Performed the work while at Meta AI as an intern.}

\IEEEauthorblockA{
University of Southern California$^{\dagger}$,
Los Angeles, USA}
\IEEEauthorblockA{
Meta AI$^{\ddagger}$, USA}
}

\maketitle

\begin{abstract}
Keyword spotting (KWS) is an important speech processing component for smart devices with voice assistance capability. 
In this paper, we investigate if Kolmogorov-Arnold Networks (KAN) can be used to enhance the performance of KWS. We explore various approaches to integrate KAN for a model architecture based on 1D Convolutional Neural Networks (CNN). We find that KAN is effective at modeling high-level features in lower-dimensional spaces, resulting in improved KWS performance when integrated appropriately. The findings shed light on understanding KAN for speech processing tasks and on other modalities for future researchers.
\end{abstract}

\section{Introduction}
\label{sec:intro}
Recent developments in voice assistance have enabled users to interact with smart devices hands-free, enriching the user experience and enhancing accessibility for physically impaired individuals. One entry point for voice assistance is Keyword Spotting (KWS), which detects specific words spoken in a streaming manner on the device. This is also referred to as wake word detection, as it helps conserve energy by only ``waking up'' the device when pre-defined keywords are detected. It is important to design an efficient KWS system that can run with minimal computational and memory consumption while minimizing false alarms and missed detections.

The advancements in deep learning have accelerated the development of Keyword Spotting (KWS) \cite{kws-dnn}. Researchers have proposed various architectures, including Convolutional Neural Networks (CNN) \cite{kws-cnn, kws-cnn2, kws-cnn3}, Recurrent Neural Networks (RNN) \cite{kws-rnn, kws-rnn2, kws-rnn3}, and Self-Attention \cite{kws-attention, kws-attention2} to improve KWS performance. These architectures are based on Multi-layer Perceptron (MLP), which consists of learnable linear layers and fixed nonlinear activation functions (e.g., ReLU).
Recently, Kolmogorov-Arnold Networks (KAN) were proposed as an alternative to MLP \cite{kan}. KAN replaces the learnable linear layers and fixed nonlinear activation in MLP with learnable nonlinear activation functions. Although KAN improves model performance and interpretability for tasks involving symbolic formula representation, as noted in \cite{kan}, its benefits for broader machine learning tasks are still ambiguous. While KAN has shown disadvantages in most domains when applied naively, as illustrated in \cite{compare}, some early works have shown advantages of KAN applied to specific domains, including medical image segmentation with U-Net \cite{ukan} and time-series forecasting \cite{kan-time}.

In this work, we explore whether KAN can be beneficial for the Keyword Spotting (KWS) task, which requires efficient modeling of intricate speech signals. KAN has demonstrated advantages in capturing complex patterns within each neuron \cite{kan}. Specifically, we use a 1D CNN-based architecture, which has shown competitive performance in KWS with a small number of parameters \cite{kws-survey}. For our experiments, we employ LiCo-Net \cite{liconet}, a simple and relatively interpretable 1D CNN-based architecture with residual connections for streaming KWS. Our work is summarized below:
\begin{itemize}
    \item To our knowledge, this is one of the earliest explorations using KAN for a speech processing task.
    \item We conduct extensive experiments to integrate KAN in a CNN-based architecture for KWS, which can guide future researchers in designing models for speech processing and other tasks.
    \item We find that KAN is effective at modeling ``high level'' complex features in low dimension, achieving improved performance for KWS through our proposed method.
\end{itemize}

\section{Kolmogorov-Arnold Networks (KAN)}
\label{sec:kan}
\subsection{KAN for DNN}
In Deep Neural Networks (DNN), KAN consists of layers of learnable nonlinear activation functions \cite{kws-dnn}. Mathematically, it can be expressed as:
$$
KAN(x) = (\Phi_{L-1} \circ \Phi_{L-2} \circ \cdot\cdot\cdot \Phi_{0})(x),
$$
where $L$ is the number of KAN layers, and $\Phi_l$ is a $n_{l+1} \times n_l$ matrix where $n_{l+1}$ and $ n_l$ signifies the number of inputs and outputs for $\Phi_l$, respectively. The entries $\phi_{l, i, j}$ for $\Phi_l$ are learnable activation functions, originally represented by B-spline basis functions with the following formula ($l, i, j$ omitted):
$$
\phi = w_b \cdot b(x) + w_s \cdot  \sum_{m} c_m \cdot B_m(x).
$$
Here, $w_b$ and $w_s$ are pre-determined weights, $b(x)$ is a basis function such as $silu(x)$, $B_m$ are B-splines, and $c_m$ are learnable parameters. 

\subsection{KAN for 1D CNN}
When KAN is applied to CNN \cite{gkan_conv, kan_conv}, the output from a 1D CNN with KAN can be expressed as follows:
$$
y_{C_{out}, i} = \sum_{d=0}^{C_{in}-1} \sum_{a=0}^{k-1} \phi_{C_{out}, a, d}(x_{d, i+a}),
$$
where $C_{in}$ and $C_{out}$ represent the input and output feature dimension sizes, respectively, and $k$ represents the kernel size. We choose Gram polynomials as the basis functions for the learnable activation functions $\phi$ instead of the original B-spline, based on our initial experiments and their superior performance for image classification tasks from \cite{gkan_conv}. We refer to KAN implemented with Gram polynomials as GKAN for the rest of the paper.

\section{Methods}
\label{sec:method}
\subsection{Keyword Spotting Pipeline}
\begin{figure}[t]
  \centering
  \centerline{\includegraphics[width=0.6\linewidth]{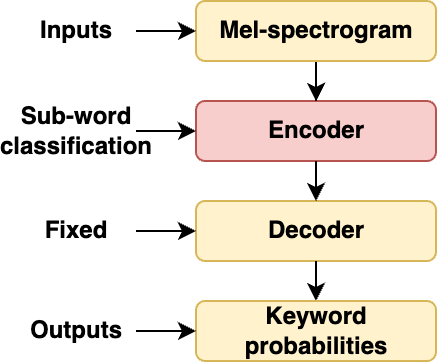}}
  \caption{Overall pipeline. We implement Encoder with LiCo-Net architecture.}\medskip
  \label{fig:pipeline}

\end{figure}

The overall KWS pipeline is illustrated in Fig~\ref{fig:pipeline}. The system takes 40-dimensional log Mel-filterbank energies as input features, computed every 10ms with a window size of 25ms. These features are fed into the Encoder, which performs the sub-word classification task. Based on the posterior probabilities of the sub-words, a Weighted Finite-State Transducers (WFST) decoder is used to estimate the probability of a predefined keyword being spoken at each timestamp.

\subsection{LiCo-Net}
\begin{figure}[t]
  \centering
  \centerline{\includegraphics[width=1\linewidth]{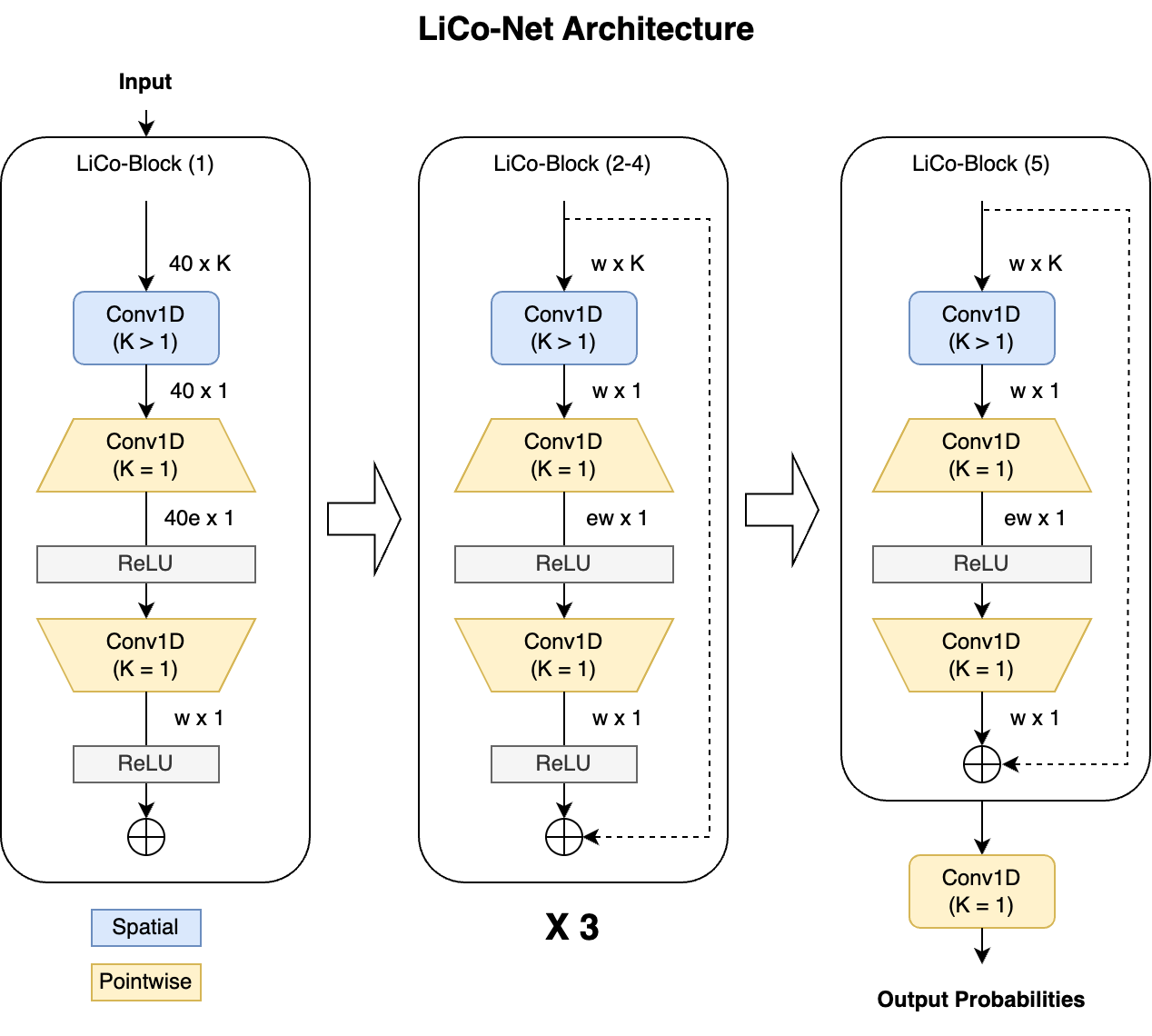}}
  \caption{Detailed illustration of LiCo-Net architecture. $K$ is kernel size, $e$ is expansion ratio, and $w$ is the dimension size.}\medskip
  \label{fig:liconet}
\end{figure}

We use Linearized Convolution Network (LiCo-Net, Fig~\ref{fig:liconet}) as the Encoder. LiCo-Net consists of sub-components named LiCo-Blocks. Each LiCo-Block consists of three 1D Convolutional (Conv1D) layers. The first Conv1D has a kernel size $K > 1$, modeling the local contexts. For convenience, we call the first Conv1D as spatial CNN. The second and third Conv1D layers are point-wise CNNs with kernel size 1. The first point-wise CNN expands the dimensional of the latent embedding by the expansion ratio $e$, allowing high dimensional modelings in the latent space. Additional Conv1D is appended at the end for the sub-word probabilities.
\label{sec:liconet}

\subsection{GKAN as Replacement Layer}
\label{sec:liconet_gkan}
\begin{figure}[t]
  \centering
  \centerline{\includegraphics[width=1\linewidth]{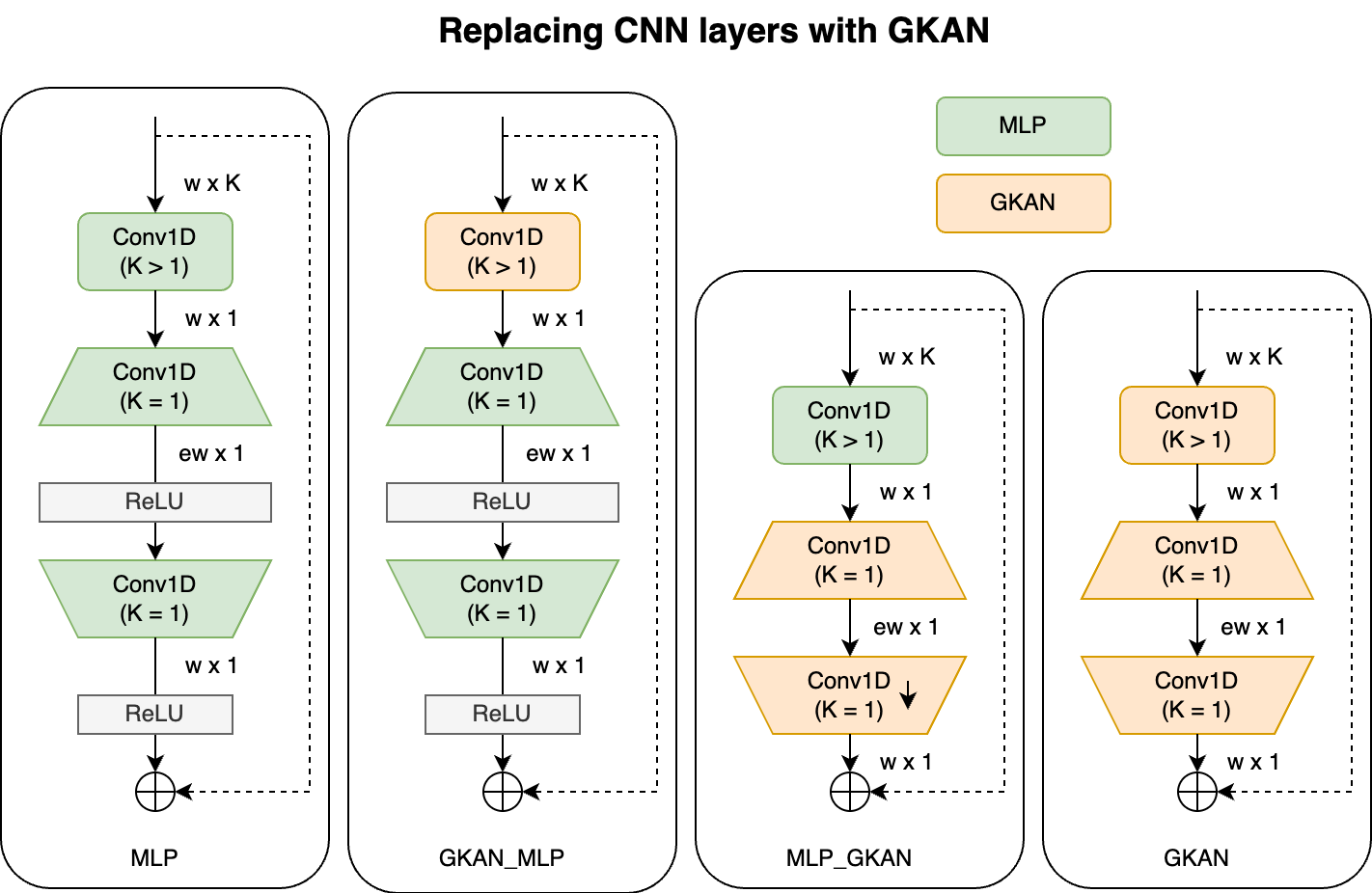}}
 
  \caption{Illustrations of LiCo-Blocks replaced with GKAN. All five LiCo-Blocks are replaced similarly for the LiCo-Net.}\medskip
  \label{fig:liconet_gkan}
  
\end{figure}

As discussed in ~\ref{sec:liconet}, a LiCo-Block consists of a spatial CNN for ``low level'' modeling with local context and a point-wise CNN for ``high level'' modeling with expanded dimensionality. To investigate whether GKAN can show advantages over MLP, we replace either or both of the spatial and point-wise CNNs with the GKAN implementation, resulting in three new architectures excluding the original one. For simplicity, we refer to these architectures as shown in Fig~\ref{fig:liconet_gkan}.

\subsection{GKAN as Additional Layer}
\begin{figure}[t]
  \centering
  \centerline{\includegraphics[width=0.9\linewidth]{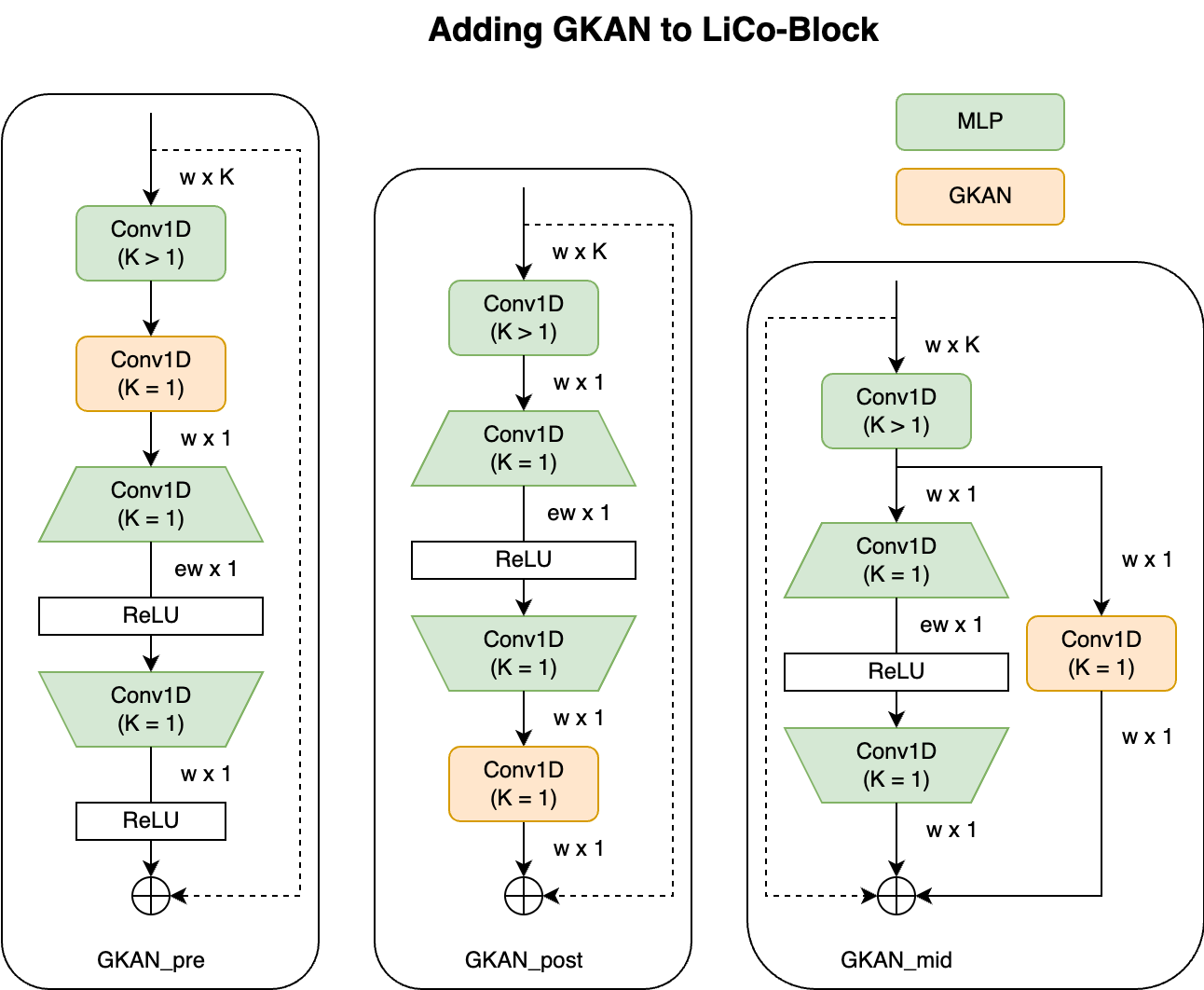}}
  \caption{Illustrations of LiCo-Blocks with an additional GKAN layer. GKAN is added similarly in all five LiCo-Blocks for the LiCo-Net.}\medskip
  \vspace{-5mm}
  \label{fig:liconet_gkan_add}
\end{figure}

While replacing Conv1Ds with GKAN as in~\ref{sec:liconet_gkan} allows us to compare the performances of MLP and KAN in a naive way, each GKAN neuron takes $O(d)$ times more parameters compared to MLP, where $d$ indicates the number of basis functions for a learnable non-linear activation function.  
Thus, we explore if GKAN can be incorporated into the architecture without substantially reducing channel width $w$ for the same number of total parameters. To this end, we insert an additional GKAN Conv1D layer where both input and output dimensions are $w \times 1$. We add a GKAN Conv1D layer before, after, and parallel to the pointwise Conv1D layers. For simplicity, we refer to these architectures as shown in Fig~\ref{fig:liconet_gkan_add}. We set the expansion ratio $e = 6$ for all of them.

\section{Experiments}
\label{sec:exp}
\subsection{Dataset}
We selected four common phrases, \textit{take a picture}, \textit{volume up}, \textit{volume down}, and \textit{play music}, as the keywords for our experiments. The aggregated and de-identified positive dataset was gathered via crowdsourcing. We divided the data into training and testing sets in a speaker-independent manner. There are around 15K, 25K, 22K, and 24K utterances from 14957 speakers for the train set and 4K, 6K, 5K, and 6K utterances from 3801 speakers for the evaluation set, respectively, with an average duration of 2.5 seconds per utterance.

We use Librispeech as the negative data not containing the keywords, where \textit{train-clean-360} and \textit{train-other-500} are used as the train set and \textit{train-clean-100} and \textit{dev/test-clean/other} are used as the evaluation set. Both the positive and the negative training sets were augmented by adding background noise and varying the speed to increase data diversity.

\subsection{Experimental Setup}
Similar to \cite{liconet}, we set the kernel size K = 5 for spatial CNN in all our experiments. The degree of Gram polynomials is set to 3. We train the models to predict 13 classes, including 11 subwords in the keyword phrases, SIL (silence), and FILLER (any other speech). We use force alignment with a pre-trained acoustic model to obtain the target labels. The 11 subwords are \textit{take}, \textit{a}, \textit{pic}, \textit{ture}, \textit{vo}, \textit{lume}, \textit{up}, \textit{down}, \textit{play}, \textit{mu}, and \textit{sic}. We use 8 GPUs with a batch size of 400, an Adam optimizer with $\beta = (0.9, 0.98)$ and a learning rate of 0.001 for 10 epochs. We use cross-entropy loss with the weights of 1, 2, and 8 for SIL, FILLER, and the 11 sub-words. For evaluation, we report false reject rates (FRR) with false accept rates per hour (FA/h) at 0.1, 0.2, 0.5, and 1.0.
\label{sec:experimetal_setup}

\section{Results and Discussion}
\subsection{GKAN as Replacement Layer}
The FRRs are shown in Table~\ref{tab:replace}. The channel width $w$ of the LiCo-Net architecture becomes much smaller for similar parameter sizes when replacing MLP Conv1D layers with GKAN Conv1D layers. We can observe that MLP\_GKAN, which uses GKAN for pointwise CNNs for high-level modeling, outperforms GKAN\_MLP and GKAN (see Fig~\ref{fig:liconet_gkan}). Thus, we can infer that GKAN is more effective at modeling high-level features, and that incorporating it in later stages of deep learning architectures could be beneficial.
\begin{table}[t]
  
  \centering
  \caption{FRR with GKAN as replacement layer. $\mathbf{w}$ is the channel width and size is the number of parameters (K).}
  \vspace{-3mm}
  \footnotesize
  \begin{tabular}{l c c c c c c}
  \toprule
  \label{tab:replace}
    \multirow{2}{*}{\textbf{Modeling}} & \multirow{2}{*}{$\mathbf{w}$} & \multirow{2}{*}{\textbf{size}} & \multicolumn{4}{c}{\textbf{FRR (\%) at FA/h =}} \\

     &  &  & \multicolumn{1}{c}{\textbf{0.1}} & \multicolumn{1}{c}{\textbf{0.2}} & \multicolumn{1}{c}{\textbf{0.5}} & \multicolumn{1}{c}{\textbf{1.0}} \\
    \cmidrule(lr){1-1} \cmidrule(lr){2-2} \cmidrule(lr){3-3} \cmidrule(lr){4-7}
    MLP (original) & $72$ & $394$ & $2.108$ & $1.718$ & $\mathbf{1.213}$ & $\mathbf{0.866}$ \\
    GKAN\_MLP & $52$ & $415$ & $2.584$ & $2.047$ & $1.429$ & $1.077$ \\
    MLP\_GKAN & $42$ & $394$ & $\mathbf{2.049}$ & $\mathbf{1.588}$ & $1.256$ & $0.923$ \\
    GKAN & $34$ & $402$ & $2.544$ & $2.077$ & $1.419$ & $1.075$ \\
    \bottomrule
  \vspace{-5mm}
  \end{tabular}
  
  \label{tab:visual}
\end{table}

\subsection{GKAN as Additional Layer}
\label{sec: gkan_add_results}
\begin{figure}[t]
  \centering
  \centerline{\includegraphics[width=0.65\linewidth]{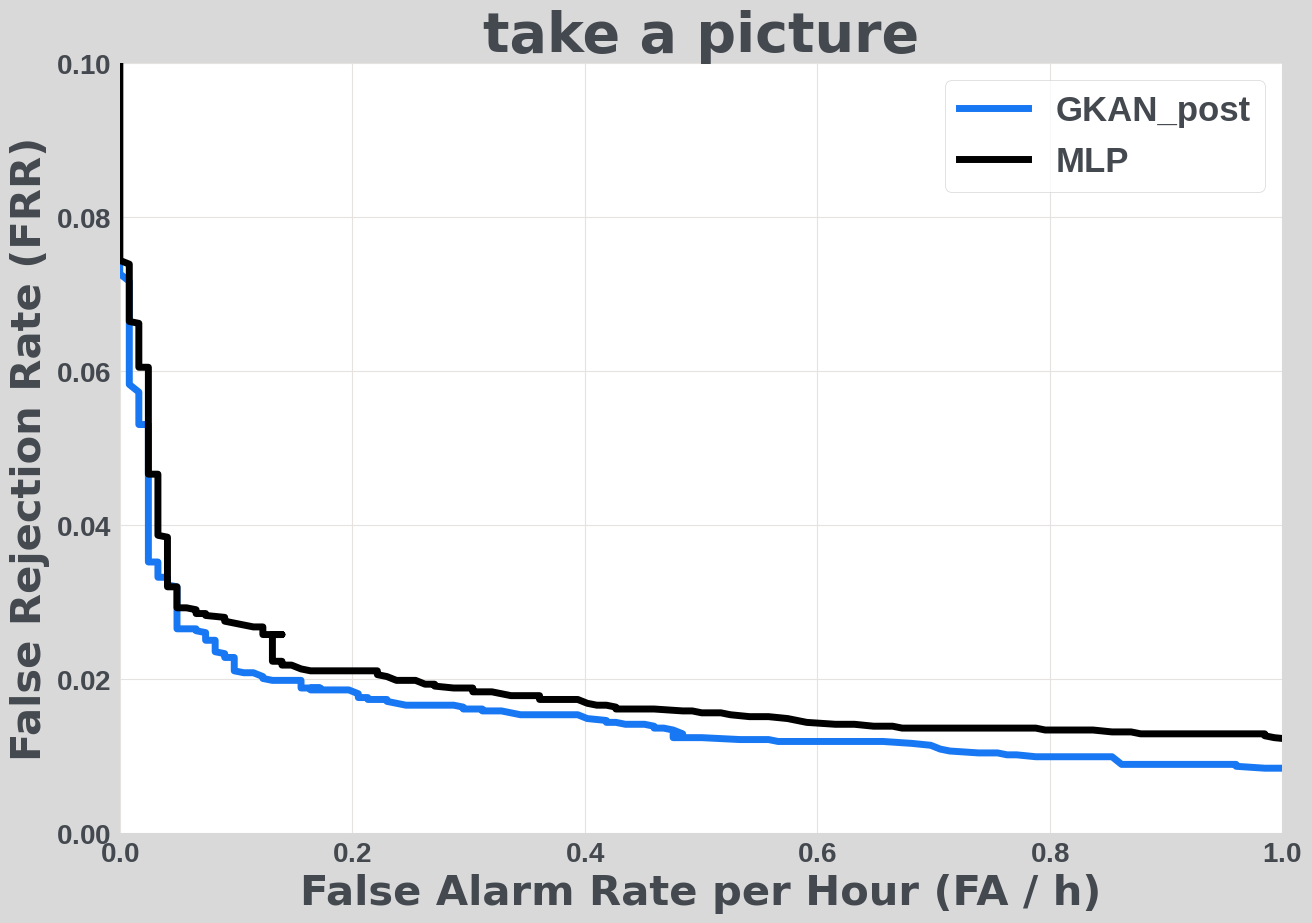}}
 \vspace{0.5mm}
  \centering
  \centerline{\includegraphics[width=0.65\linewidth]{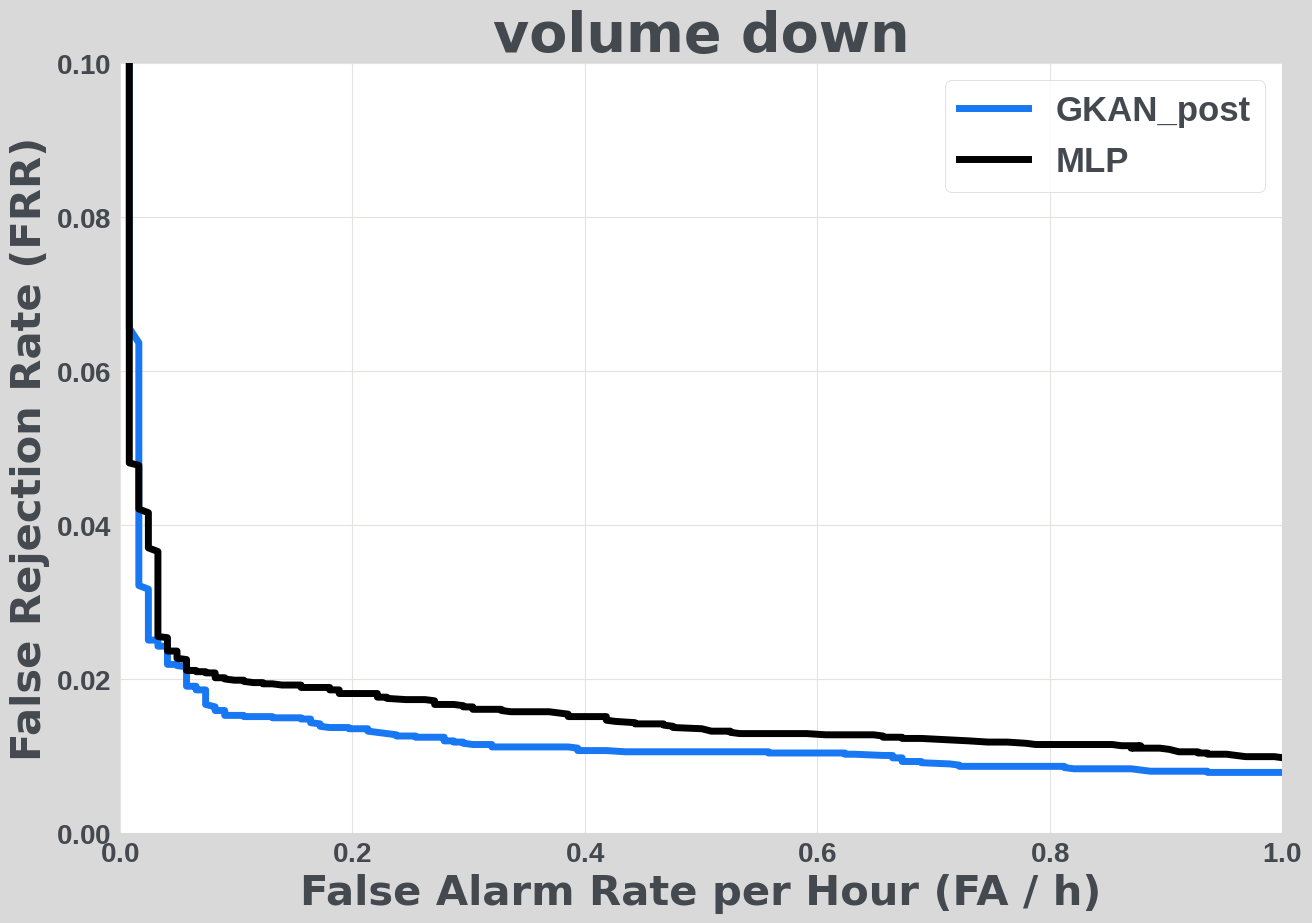}}
  \caption{DET curves for \textit{take a picture} and \textit{volume down}.}\medskip
  \label{fig:det}
  \vspace{-5mm}
\end{figure}

The FRRs are shown in Table~\ref{tab:adding}. The decrease in LiCo-Net channel width $w$ is less substantial compared to Sec~\ref{sec:liconet_gkan}. GKAN\_post, which introduces a GKAN Conv1D layer after the pointwise CNNs for the high-level modeling within LiCo-Block as shown in Fig~\ref{fig:liconet_gkan_add}, results in the lowest FRRs compared to other architectures, including the original LiCo-Net. DET curves for GKAN\_post with two keywords, \textit{take a picture} and \textit{volume down}, are shown in Fig~\ref{fig:det}. Again, we observe that GKAN is better suited for modeling high-level features as discussed in Sec~\ref{sec:liconet_gkan}. We also see that GKAN\_mid is substantially underperforming other implementations, possibly because the differences in how MLP and GKAN model input features make them unsuitable to be used in parallel. In addition, we observe that GKAN's advantages are more pronounced when implemented in lower dimensions, as it can be incorporated without significantly increasing the number of parameters. Intuitively, GKAN or any implementations of KAN lead to several times more parameters per neuron, making the network ``dense''. Thus, it is likely to show advantages when modeling complex patterns in low-dimensional features.

\begin{table}[t]
  
  \centering
  \caption{FRR after adding GKAN. $\mathbf{w}$ is the channel width and size is the number of parameters (K).}
  \vspace{-3mm}
  \footnotesize
  \begin{tabular}{l c c c c c c}
  \toprule
  \label{tab:adding}
    \multirow{2}{*}{\textbf{Modeling}} & \multirow{2}{*}{$\mathbf{w}$} & \multirow{2}{*}{\textbf{size}} & \multicolumn{4}{c}{\textbf{FRR (\%) at FA/h =}} \\

     &  &  & \multicolumn{1}{c}{\textbf{0.1}} & \multicolumn{1}{c}{\textbf{0.2}} & \multicolumn{1}{c}{\textbf{0.5}} & \multicolumn{1}{c}{\textbf{1.0}} \\
    \cmidrule(lr){1-1} \cmidrule(lr){2-2} \cmidrule(lr){3-3} \cmidrule(lr){4-7}
    MLP (original) & $72$ & $394$ & $2.108$ & $1.718$ & $1.213$ & $0.866$ \\
    GKAN\_pre & $64$ & $408$ & $2.226$ & $1.863$ & $1.272$ & $0.974$ \\
    GKAN\_post & $62$ & $396$ & $\mathbf{1.772}$ & $\mathbf{1.525}$ & $\mathbf{1.073}$ & $\mathbf{0.758}$ \\
    GKAN\_mid & $64$ & $412$ & $3.446$ & $2.537$ & $1.939$ & $1.444$ \\
    \bottomrule
  \vspace{-5mm}
  \end{tabular}
  
  \label{tab:visual}
\end{table}

\subsection{Robustness to Noisy Conditions}
To evaluate if GKAN\_post is robust in noisier conditions, we have augmented the evaluation dataset by adding randomly sampled background noise, which encompasses a wide range of categories. The SNR level of the augmented set follows a Gaussian distribution centered at 12.5db and bounded by [0, 25] db. The FRRs are shown in Table~\ref{tab:noisy}. GKAN\_post is outperforming MLP in all the FA/h thresholds, showing its robustness in noisier situations.

\begin{table}[t]
  \centering
  \caption{FRR with noisy dataset. $\mathbf{w}$ is the channel width and size is the number of parameters (K).}
  \vspace{-3mm}
  \footnotesize
  \begin{tabular}{l c c c c c c}
    \toprule
    \label{tab:noisy}
    \multirow{2}{*}{\textbf{Modeling}} & \multirow{2}{*}{$\mathbf{w}$} & \multirow{2}{*}{\textbf{size}} & \multicolumn{4}{c}{\textbf{FRR (\%) at FA/h =}} \\

     &  &  & \multicolumn{1}{c}{\textbf{0.1}} & \multicolumn{1}{c}{\textbf{0.2}} & \multicolumn{1}{c}{\textbf{0.5}} & \multicolumn{1}{c}{\textbf{1.0}} \\
    \cmidrule(lr){1-1} \cmidrule(lr){2-2} \cmidrule(lr){3-3} \cmidrule(lr){4-7}
    MLP (original) & $72$ & $394$ & $8.748$ & $7.017$ & $4.903$ & $4.024$ \\
    GKAN\_post & $62$ & $396$ & $\mathbf{7.748}$ & $\mathbf{6.167}$ & $\mathbf{4.332}$ & $\mathbf{3.434}$ \\
    \bottomrule
  \vspace{-5mm}
  \end{tabular}
  \label{tab:visual}
\end{table}

\begin{table}[t]
  \centering
  \caption{FRR varying the model sizes. $\mathbf{w}$ is the channel width and size is the number of parameters (K).}
  \vspace{-3mm}
  \footnotesize
  \begin{tabular}{l c c c c c c}
    \toprule
    \label{tab:sizes}
    \multirow{2}{*}{\textbf{Modeling}} & \multirow{2}{*}{$\mathbf{w}$/\textbf{size}} & \multirow{2}{*}{\textbf{data}} & \multicolumn{4}{c}{\textbf{FRR (\%) at FA/h =}} \\

     &  &  & \multicolumn{1}{c}{\textbf{0.1}} & \multicolumn{1}{c}{\textbf{0.2}} & \multicolumn{1}{c}{\textbf{0.5}} & \multicolumn{1}{c}{\textbf{1.0}} \\
    \cmidrule(lr){1-1} \cmidrule(lr){2-2} \cmidrule(lr){3-3} \cmidrule(lr){4-7}
    \cmidrule(lr){1-1} \cmidrule(lr){2-2} \cmidrule(lr){3-3} \cmidrule(lr){4-7}
    MLP & $50/204$ & clean & $2.55$ & $2.04$ & $\mathbf{1.39}$ & $1.09$ \\
    GKAN\_post & $42/196$ & clean & $\mathbf{2.16}$ & $\mathbf{1.83}$ & $1.41$ & $\mathbf{0.97}$ \\
    \cmidrule(lr){1-1} \cmidrule(lr){2-2} \cmidrule(lr){3-3} \cmidrule(lr){4-7}
    MLP & $36/104$ & clean & $\mathbf{3.23}$ & $2.72$ & $1.99$ & $1.56$ \\
    GKAN\_post & $30/101$ & clean & $3.64$ & $\mathbf{2.68}$ & $\mathbf{1.87}$ & $\mathbf{1.44}$ \\
    \cmidrule(lr){1-1} \cmidrule(lr){2-2} \cmidrule(lr){3-3} \cmidrule(lr){4-7}
    MLP & $50/204$ & noisy & $10.51$ & $8.46$ & $6.18$ & $4.82$ \\
    GKAN\_post & $42/196$ & noisy & $\mathbf{10.13}$ & $\mathbf{7.74}$ & $\mathbf{5.44}$ & $\mathbf{4.25}$ \\
    \cmidrule(lr){1-1} \cmidrule(lr){2-2} \cmidrule(lr){3-3} \cmidrule(lr){4-7}
    MLP & $36/104$ & noisy & $\mathbf{12.56}$ & $\mathbf{10.12}$ & $7.68$ & $6.17$ \\
    GKAN\_post & $30/101$ & noisy & $12.94$ & $10.27$ & $\mathbf{7.58}$ & $\mathbf{5.94}$ \\
    \bottomrule
  \end{tabular}
  
  \label{tab:visual}
\end{table}

\subsection{Analysis on Smaller Model Sizes}
To assess the effectiveness of GKAN\_post for smaller model sizes,  we evaluate the models with approximately 100K and 200K parameters. We use an expansion ratio $e$ of 5 and 6 for the 100K and 200K models, respectively, with varying channel width $w$, while all other hyper-parameters are kept the same. The FRR scores are shown in Table~\ref{tab:sizes}. While GKAN\_post with 200K parameter size outperforms the MLP implementation overall, GKAN\_post with 100K parameter size shows comparable results to MLP. This is likely because the decrease in channel width $w$ is more detrimental for smaller models. 

\label{sec:results}

\section{ablation study}
\begin{figure}[t]
  \centering
  \centerline{\includegraphics[width=0.6\linewidth]{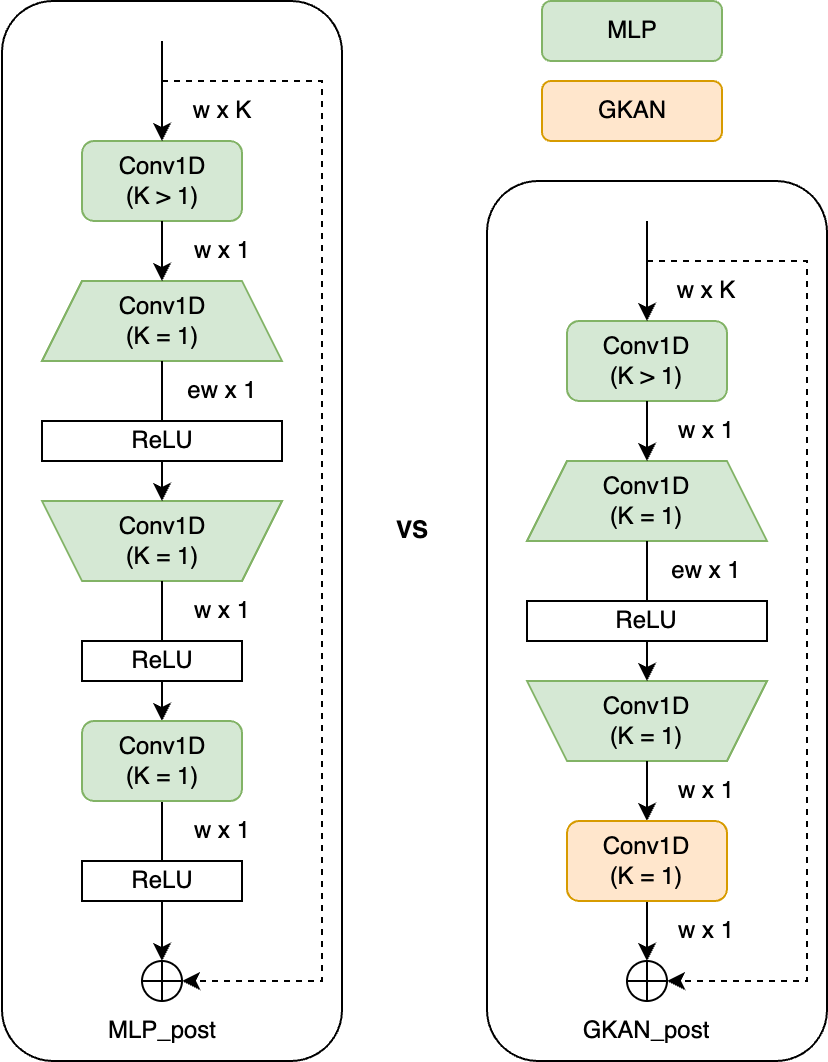}}
  \caption{Ablation study. Additional Conv1D layer is added to the original LiCo-Block with MLP, to match the structure of GKAN\_post.}\medskip
  \vspace{-5mm}
  \label{fig:liconet_ablation}
\end{figure}

The comparison between the original LiCo-Net and GKAN\_post may not be completely fair, since adding GKAN Conv1D changes the structure of the LiCo-Net architecture. Thus, we have added a Conv1D layer to the original LiCo-Block (referred as MLP\_post), to match the GKAN\_post architecture, as shown in Figure~\ref{fig:liconet_ablation}.
We use MLP\_post with 399K total number of parameters and the channel width $w$ of 70 to compare against the GKAN\_post model as in Sec~\ref{sec: gkan_add_results}.  As in Table~\ref{tab:ablation}, we can see the strong performance of GKAN\_post over MLP\_post for both clean and noisy evaluations, confirming the benefits of the proposed GKAN integration method.

\begin{table}[!t]
  \centering
  \caption{Comparing MLP\_post and GKAN\_post.}
  \footnotesize
  \begin{tabular}{l c c c c c}
    \toprule
    \label{tab:ablation}
    \multirow{2}{*}{\textbf{Modeling}} & \multirow{2}{*}{\textbf{data}} & \multicolumn{4}{c}{\textbf{FRR (\%) at FA/h =}} \\

     &  &  \multicolumn{1}{c}{\textbf{0.1}} & \multicolumn{1}{c}{\textbf{0.2}} & \multicolumn{1}{c}{\textbf{0.5}} & \multicolumn{1}{c}{\textbf{1.0}} \\
    \cmidrule(lr){1-1} \cmidrule(lr){2-2} \cmidrule(lr){3-6}
    MLP\_post & clean & $2.06$ & $1.658$ & $1.12$ & $0.832$ \\
    GKAN\_post & clean& $\mathbf{1.772}$ & $\mathbf{1.525}$ & $\mathbf{1.073}$ & $\mathbf{0.758}$ \\
    \cmidrule(lr){1-1} \cmidrule(lr){2-2} \cmidrule(lr){3-6}
    MLP\_post & noisy & $9.3$ & $7.118$ & $5.124$ & $3.943$ \\
    GKAN\_post & noisy & $\mathbf{7.748}$ & $\mathbf{6.167}$ & $\mathbf{4.332}$ & $\mathbf{3.434}$ \\
    \bottomrule
  \vspace{-5mm}
  \end{tabular}
  
  \label{tab:visual}
\end{table}

\section{Conclusion}
In this work, we systematically study how Kolmogorov-Arnold Networks (KAN) can be effectively integrated into KWS modeling with a Conv1D-based architecture. Through extensive experimentation, we find that Gram KAN (GKAN) is most effective when modeling high-level features in low-dimensional space. Our proposed method, GKAN\_post, significantly reduces FRR when used with appropriate model parameter sizes. While this work focuses on KWS in speech processing,  the findings can facilitate future researchers to effectively integrate KAN in other domains.

\label{sec:conclusion}


  


\begin{thebibliography}{00}

\bibitem{kws-dnn} G. Chen, C. Parada, and G. Heigold. ``Small-footprint keyword spotting using deep neural networks,'' IEEE international conference on acoustics, speech and signal processing (ICASSP) (pp. 4087-4091), IEEE, 2014.
\bibitem{kws-cnn} R. Tang and J. Lin. ``Deep residual learning for small-footprint keyword spotting,'' IEEE International Conference on Acoustics, Speech and Signal Processing (ICASSP) (pp. 5484-5488), IEEE, 2018.
\bibitem{kws-cnn2} Y. Zhang, N. Suda, L. Lai, and V. Chandra. ``Hello edge: Keyword spotting on microcontrollers,'' arXiv preprint arXiv:1711.07128, 2017.
\bibitem{kws-cnn3} M. Sun, et al. `Compressed time delay neural network for small-footprint keyword spotting,'' Interspeech, 2017.
\bibitem{kws-rnn} S. Fernández, A. Graves, and J. Schmidhuber. ``An application of recurrent neural networks to discriminative keyword spotting.'' In International conference on artificial neural networks (pp. 220-229), Springer Berlin Heidelberg, 2007.
\bibitem{kws-rnn2} M. Sun, et al. ``Max-pooling loss training of long short-term memory networks for small-footprint keyword spotting,'' IEEE spoken language technology workshop (SLT) (pp. 474-480). IEEE, 2016.
\bibitem{kws-rnn3} S. O. Arik, et al. ``Convolutional recurrent neural networks for small-footprint keyword spotting,'' arXiv preprint arXiv:1703.05390, 2017.

\bibitem{kws-attention} A. Berg, M. O'Connor, and M. T. Cruz (2021). ``Keyword transformer: A self-attention model for keyword spotting,'' Interspeech, 2021.
\bibitem{kws-attention2} O. Rybakov, N. Kononenko, N. Subrahmanya, M. Visontai, and S. Laurenzo. (2020). ``Streaming keyword spotting on mobile devices,'' Interspeech, 2021.
\bibitem{kan} Z. Liu, et al. ``Kan: Kolmogorov-arnold networks,'' arXiv preprint arXiv:2404.19756 , 2024.
\bibitem{compare} R. Yu, W. Yu, and X. Wang. ``Kan or mlp: A fairer comparison,'' arXiv preprint arXiv:2407.16674, 2024
\bibitem{ukan} C. Li, X. Liu, W. Li, C. Wang, H. Liu, and Y Yuan. ``U-KAN Makes Strong Backbone for Medical Image Segmentation and Generation,'' arXiv preprint arXiv:2406.02918, 2024.
\bibitem{kan-time} C. J. Vaca-Rubio, L. Blanco, R. Pereira, and M. Caus. ``Kolmogorov-arnold networks (kans) for time series analysis,'' IEEE International Workshop on Machine Llearning for Signal Processing, 2024.
\bibitem{kws-survey} I. López-Espejo, Z. H. Tan, J. H. Hansen, and H. Jensen. ``Deep spoken keyword spotting: An overview,'' IEEE Access, 10, 4169-4199, 2021.
\bibitem{liconet} H. Yang, et al. ``Lico-net: Linearized convolution network for hardware-efficient keyword spotting,'' arXiv preprint arXiv:2211.04635, 2022.
\bibitem{kan_conv} A. D. Bodner, A. S. Tepsich, J. N. Spolski., and S. Pourteau. Convolutional Kolmogorov-Arnold Networks. arXiv preprint arXiv:2406.13155, 2024.
\bibitem{gkan_conv} I. Drokin. Kolmogorov-Arnold ``Convolutions: Design Principles and Empirical Studies,'' arXiv preprint arXiv:2407.01092, 2024.
\bibitem{kaldi} D. Povey, et al. ``The Kaldi Speech Recognition Toolkit,'' Workshop on Automatic Speech Recognition and Understanding, IEEE, 2011.

\end{thebibliography}
\end{document}